\documentclass[
    ,final            % use final for the camera ready runs
  ]
  {aipproc}

\layoutstyle{6x9}
\def\tev{\ifmmode \mathop{\rm TeV}\nolimits\else {\rm TeV}\fi}
\def\gev{\ifmmode \mathop{\rm GeV}\nolimits\else {\rm GeV}\fi}
\def\mev{\ifmmode \mathop{\rm MeV}\nolimits\else {\rm MeV}\fi}
\def\kev{\ifmmode \mathop{\rm keV}\nolimits\else {\rm keV}\fi}
\def\ev{\ifmmode \mathop{\rm eV}\nolimits\else {\rm eV}\fi}
\def\ryd{\ifmmode \mathop{\rm Ry}\nolimits\else {\rm Ry}\fi}
\def\angst{\ifmmode\mathop{\rm\AA}\nolimits\else {\rm \AA}\fi}

\def\pepe{\mathop{\rm P.P.}}

\def\real{\mathop{\rm Re}}
\def\imag{\mathop{\rm Im}}

\def\dd{{\rm d}}

\def\degreec{\ifmmode\mathop{^\circ \rm C}\nolimits\else{$^\circ{\rm C}\;$}\fi}
\def\degreek{\ifmmode\mathop{^\circ \rm K}\nolimits\else{$^\circ{\rm K}\;$}\fi}
\def\degreef{\ifmmode\mathop{^\circ \rm F}\nolimits\else{$^\circ{\rm F}\;$}\fi}

\def\chidof{\ifmmode\mathop\chi^2/{\rm d.o.f.}\else $\chi^2/{\rm d.o.f.}\null$\fi}

\def\msbar{\ifmmode\mathop{\overline{\rm MS}}\else$\overline{\rm MS}$\null\fi}

\def\cmass{\ifmmode\mathop{\rm c.m.}\nolimits\else {\sl c.m.}\fi}
\def\lab{\ifmmode{\rm lab}\else {\sl lab.}\fi}

\def\degrees{\ifmmode{^\circ\,}\else $^\circ$\fi}
\def\feet{\ifmmode{\hbox{'}\,}\else '\fi}
\def\inches{\ifmmode{\hbox{"}\,}\else "\fi}

\def\lsim{\mathop{\setbox0=\hbox{$\displaystyle 
\raise2.2pt\hbox{$\;<$}\kern-7.7pt\lower2.6pt\hbox{$\sim$}\;$}
\box0}}
\def\gsim{\mathop{\setbox0=\hbox{$\displaystyle 
\raise2.2pt\hbox{$\;>$}\kern-7.7pt\lower2.6pt\hbox{$\sim$}\;$}
\box0}}

\def\frac#1#2{{#1\over#2}}
\def\dfrac#1#2{{\displaystyle{#1\over#2}}}
\def\tfrac#1#2{{\textstyle{#1\over#2}}}
\def\ffrac#1#2{\leavevmode
   \kern.1em \raise .5ex \hbox{\the\scriptfont0 #1}%
   \kern-.1em $/$%
   \kern-.15em \lower .25ex \hbox{\the\scriptfont0 #2}%
}%

\begin{document}
\title{Consistency checks of pion-pion scattering data
and chiral dispersive calculations}

\author{ J. R. Pel\'aez}{
  address={Departamento de F\'{\i}sica Te\'orica,~II,
Facultad de Ciencias F\'{\i}sicas,
Universidad Complutense de Madrid,
E-28040, Madrid, Spain}
}

\author{F. J. 
Yndur\'ain\footnotetext{Presented by F. J. Yndur\'ain at 
``Quark confinement and the Hadron Spectrum", Sardinia, Sept. 2004.}}{
  address={Departamento de F\'{\i}sica Te\'orica, C-XI
 Universidad Aut\'onoma de Madrid,
 Canto Blanco,
E-28049, Madrid, Spain.}
}
\begin{abstract}
We have evaluated forward dispersion relations for 
scattering amplitudes that follow from direct
fits to several sets of $\pi\pi$
scattering experiments, together with the precise
K decay results, and high to energy data.
We find that
some of the most commonly used experimental sets,
as well as some recent theoretical analyses based on Roy equations,
do not satisfy these constraints by several standard deviations. 
Finally, we provide a consistent
$\pi\pi$ amplitude by improving a global fit
to data with these dispersion relations.
\end{abstract}
\maketitle

%%%%%%%%%%%%%%%%%%%%%%%%%%%%%%%%%%%%%%%%%%%%
%% MAINMATTER
%%%%%%%%%%%%%%%%%%%%%%%%%%%%%%%%%%%%%%%%%%%%

\section{Introduction}
A precise knowledge of the $\pi\pi$ scattering amplitude
has become increasingly important since it provides crucial tests
for one and two loop Chiral Perturbation Theory (ChPT), as well as crucial information on
three topics under intensive experimental and theoretical investigation:
light meson spectroscopy, pionic atom decays and CP violation in kaons.
Unfortunately, these precision studies are very cumbersome due
to the poor quality of the data which is affected by large systematic errors.
Here we review our recent work where we checked 
the fulfillment of dispersion relations by
different sets of data commonly used in the literature, and provided 
parametrizations consistent with such requirements.

Recently, Ananthanarayan, Colangelo, Gasser and Leutwyler (ACGL)\cite{1} 
and Colangelo, Gasser and Leutwyler (CGL)\cite{2} have used 
data, analyticity and unitarity through Roy equations, and ChPT,
to build a $\pi\pi$ amplitude. They provide phase shifts
up to $0.8\,\gev$, 
scattering lengths and effective ranges
claiming an outstanding precision.
While the methods of CGL constitute a substantial improvement over previous ones, 
their analysis has to rely on some input, part of which we have recently questioned.
First of all, their Regge high energy representation
does not describe the high energy data, and does not satisfy well
certain sum rules.
Second, their D2 wave is incompatible with a number of requirements.
Finally some of their input has remarkably small errors 
and relies precisely on some data sets that do not satisfy well forward
dispersion relations. All this is discussed in the present note,
 which is based our recent works \cite{3,4,5}.

\section{Low Energy Partial waves from fits to
data}

\subsection{The S0, S2 and P partial waves at low energy, $s^{1/2}\lsim 1\,\gev$}

We 
first consider wave-by-wave {\it fits to data} for the S0, S2, P waves, as 
in \cite{5}, which improve our ``tentative solution" in \cite{3}. 
To fit the phase shifts, $\delta(s)$, we parametrize
 $\cot\delta(s)$ taking into account its analytic properties, 
as well as its zeros (associated with resonances) 
and poles (when the phase shift crosses $n\pi$, $n=$integer).
  
For the P wave, up to $\simeq1\,$GeV we use the results from a
 fit to the pion form factor as given in \cite{6}. The comparison with $\pi\pi$
scattering data can be seen in Fig.1.
We take $s_0$ as the point at which inelasticity begins to be nonegligible,
and we write
\begin{eqnarray}
&&\cot\delta_1(s)=\dfrac{s^{1/2}}{2k^3}
(M^2_\rho-s)\left\{B_0+B_1\dfrac{\sqrt{s}-\sqrt{s_0-s}}{\sqrt{s}+\sqrt{s_0-s}}
\right\};\quad s_0^{1/2}=1.05\;\gev.
\label{(2.1a)}\\
&&B_0=\,1.069\pm0.011,\quad B_1=0.13\pm0.05,\quad M_{\rho}=773.6\pm0.9,\cr 
&&a_1=\,(37.6\pm1.1)\times10^{-3}M_{\pi}^{-3},\quad
  b_1=(4.73\pm0.26)\times10^{-3}M_{\pi}^{-5};\label{eq:2.1}
\end{eqnarray}

\begin{figure}
  \includegraphics[height=.4\textheight,angle=-90]{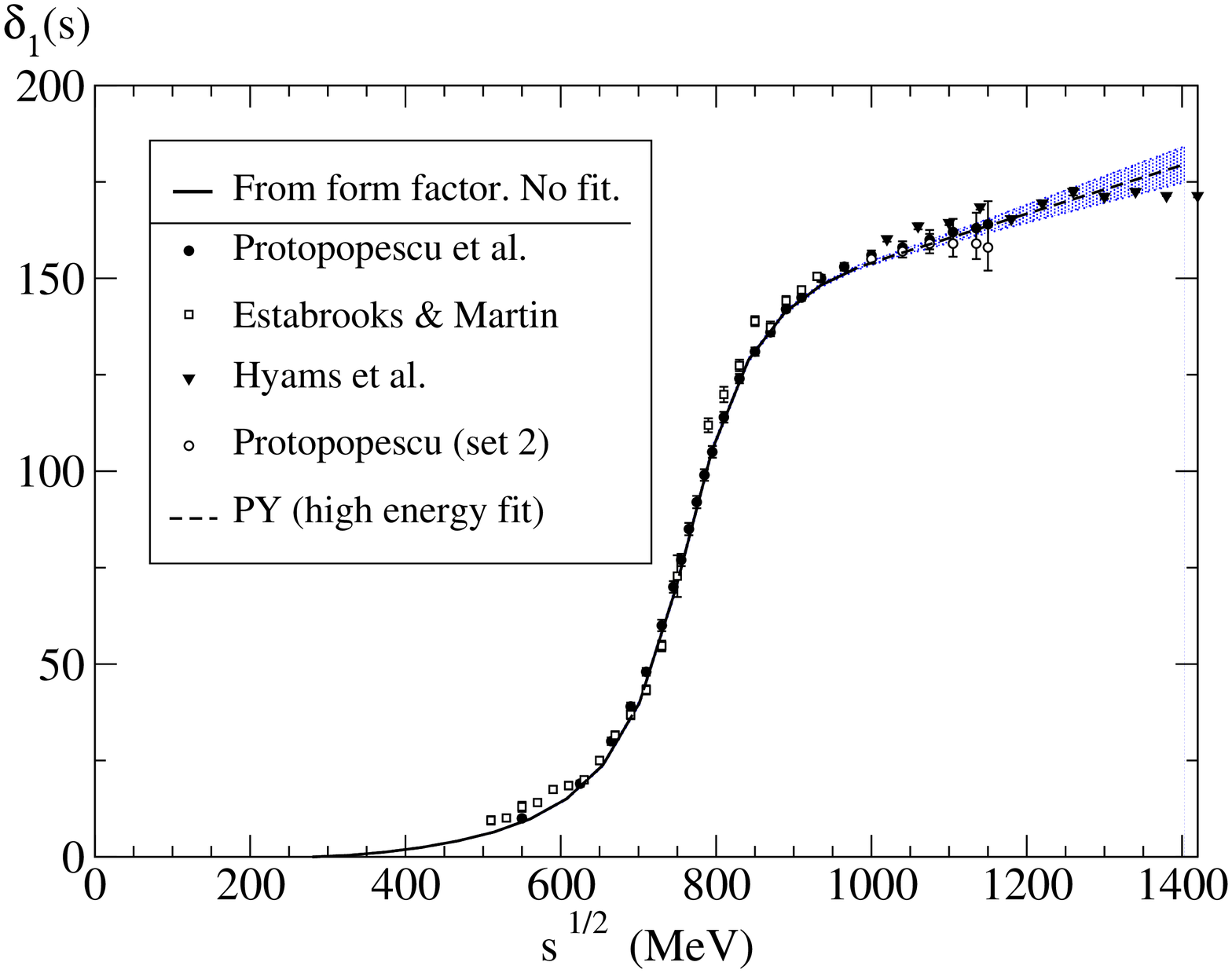}
\hspace*{-1.1cm}  
\includegraphics[height=.4\textheight,angle=-90]{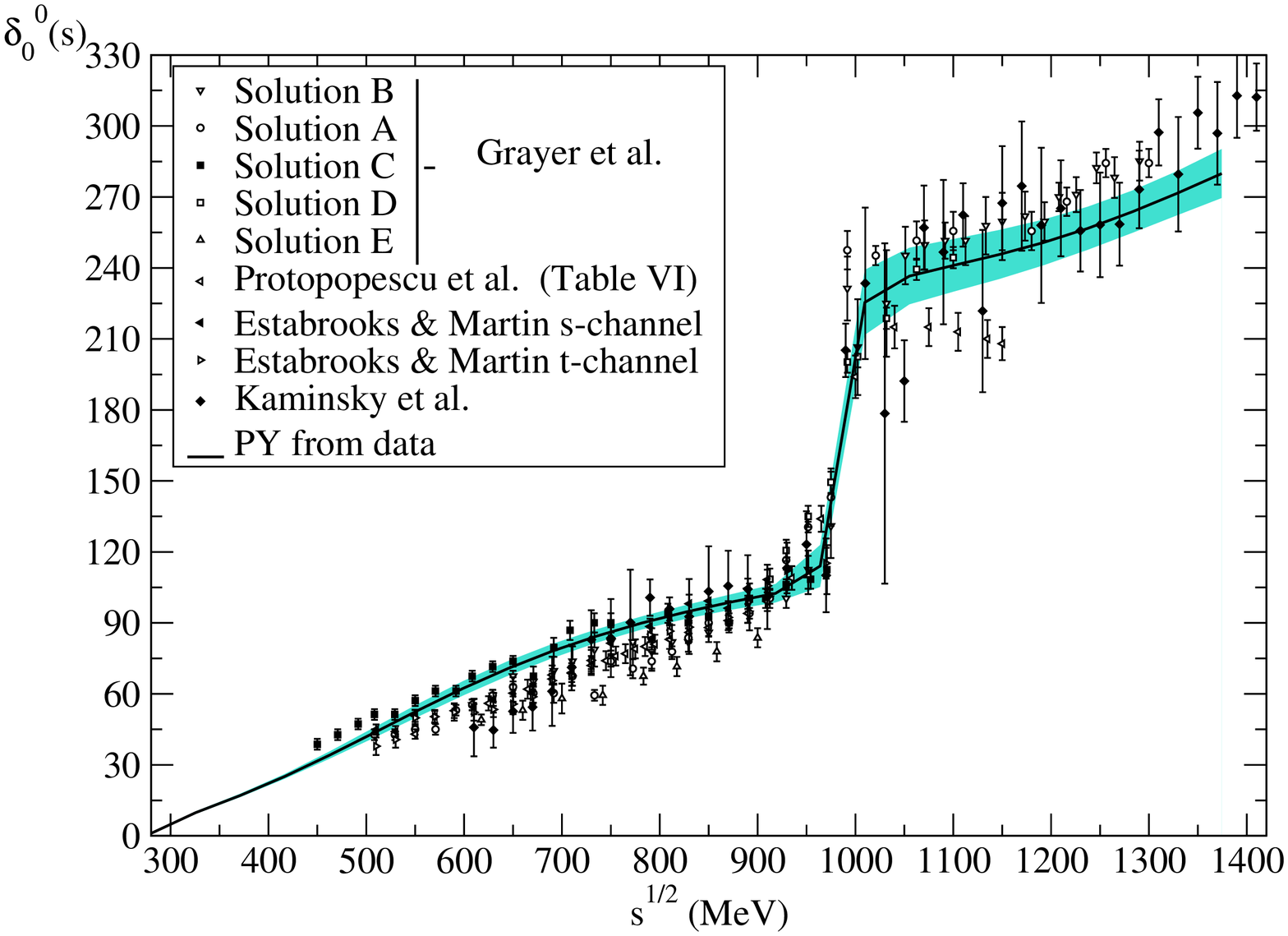}
  \caption{a) P wave phase shifts from $\pi\pi$ scattering \cite{9,10}
compared with the prediction with 
the parameters (2.1) (solid line below 1~\gev). 
Note that this is {\sl not} a fit to these data, but is obtained 
from the pion form factor \cite{6}. The error here 
is like the thickness of the line.
Above 1~\gev, the dotted line and error (PY) are as 
follows from the fit in \cite{5}. b)  
S0 phase shifts and error band as 
given by Eqs.(\ref{(2.3a)},\ref{(2.3b)}) below 1~\gev, and from \cite{5} above.  
The $K_{l4}$ and $K_{2\pi}$ decay data are not shown. (see our \cite{5}
for details).}
\end{figure}

For the S2 wave we fit data where two like charge pions are produced:\cite{7} 
although these pions are not all on their mass shell, at least there is no 
problem of interference among various isospin states. 
At low energies, we  fix the Adler zero 
at $z_2=M_\pi$ and fit only the low energy data, 
$s^{1/2}<1.0\,\gev$; later on we allow $z_2$ to vary. We 
have
\begin{eqnarray}
&&\cot\delta_0^{(2)}(s)=\dfrac{s^{1/2}}{2k}\,\dfrac{M_{\pi}^2}{s-2z_2^2}\,
\left\{B_0+B_1\dfrac{\sqrt{s}-\sqrt{s{_0}-s}}{\sqrt{s}+\sqrt{s{_0}-s}}\right\},
\quad s{_0}^{1/2}=1.05\;\gev,\\
&&B_0=\,-80.4\pm2.8,\quad B_1=-73.6\pm12.6;\cr
&&a_0^{(2)}=\,(-0.052\pm0.012)\,M_{\pi}^{-1};\quad
b_0^{(2)}=(-0.085\pm0.011)\,M_{\pi}^{-3}.
\label{(2.2b)}
\end{eqnarray}

The S0 wave experimental situation is somewhat 
confusing, and we consider two methods of data selection. 
In both, we fit $K_{l4}$ and $K\to2\pi$ decay
data, in which  pions are on the {\sl mass shell}. In the first method, 
called {\it global fit}, we
 include some points at $0.81\,\gev\leq s^{1/2}\leq0.97\,\gev$,
 where the various experiments agree within 
$\lsim1.5\,\sigma$.
Care is exercised to compose errors realistically,
see details in 
\cite{5}, Subsect~2.2.2. 
In this case we fix the Adler zero at $z_0=M_\pi$ and find
\begin{eqnarray}
&&\cot\delta_0^{(0)}(s)=\,\dfrac{s^{1/2}}{2k}\,\dfrac{M_{\pi}^2}{s-\tfrac{1}{2}z_0^2}\,
\dfrac{M^2_\sigma-s}{M^2_\sigma}\,
\left\{B_0+B_1\dfrac{\sqrt{s}-\sqrt{s_0-s}}{\sqrt{s}+\sqrt{s_0-s}}\right\},
\label{(2.3a)} \\
&&{B}_0=\,21.04,\quad {B}_1=6.62,\quad
M_\sigma=782\pm24\,\mev;\quad\delta_0^{(0)}(m_K)= 41.0\degrees\pm2.1\degrees;
\nonumber\\
&&a_0^{(0)}=\,(0.230\pm0.010) M_{\pi}^{-1},\quad b_0^{(0)}=(0.268\pm0.011)
M_{\pi}^{-3}; \nonumber 
\end{eqnarray}
this fit (shown in Fig~1.b as PY) is valid for $s^{1/2}\leq0.95\,\gev$. 
The $B_i$ errors are strongly correlated; uncorrelated errors are obtained if 
using the parameters $x,\,y$ with
\begin{equation}
B_0=y-x;\quad B_1=6.62-2.59 x;\quad y=21.04\pm0.70,\quad x=0\pm 2.6.
\label{(2.3b)}
\end{equation}

The other method is to fit only  $K_{l4}$ and $K\to2\pi$ data, or 
to add to this, individually, data from the various experimental analyses.
The results can be found in Table~1.

\subsection{ The S0, S2 and P partial waves at $1\,\gev\lsim s^{1/2}\lsim 1.42\,\gev$}
The D and F data are scanty, and have large errors. 
To stabilize the fits we impose the 
values of the scattering lengths that follow 
from the Froissart--Gribov representation. 
This is not circular reasoning since
their  Froissart--Gribov representation depends mostly on the 
S0, S2 and P waves, and very little on the
 D0, D2, F waves themselves. 
We do not discuss here the D0 and F waves (see \cite{5}) 
as they do not present special features.

For D2 we only expect important inelasticity when the 
 $\pi\pi\to\rho\rho$ channel  opens up, 
so that $s_0=1.45^2\,\gev^2\sim4M^2_\rho$.
A pole term is necessary here,
since we expect $\delta_2^{(2)}$ to change sign near threshold:
the data \cite{7} give 
negative and small values for $\delta^{(2)}_2$ above some $500\,\mev$, while, 
from  the Froissart--Gribov representation, 
it is known\cite{12} that the  
scattering length must be positive. Indeed we include in the fit
the value
$a_2^{(2)}=(2.72\pm0.36)\times10^{-4}\,M_{\pi}^{-5}$.
In addition, the clear inflection seen in data around 1~\gev\ 
asks for a third order conformal expansion. So 
we write
\begin{eqnarray}
\cot\delta_2^{(2)}(s)=
\dfrac{s^{1/2}}{2k^5}\,\Big\{B_0+B_1 w(s)+B_2 w(s)^2\Big\}\,
\dfrac{{M_\pi}^4 s}{4({M_\pi}^2+\Delta^2)-s},\;
w(s)=\dfrac{\sqrt{s}-\sqrt{s_0-s}}{\sqrt{s}+\sqrt{s_0-s}}.\nonumber
\label{(2.4a)}
\end{eqnarray}
And we find
$B_0=(2.4\pm0.3)\times10^3$, $B_1=(7.8\pm0.8)\times10^3$,
 $B_2=(23.7\pm3.8)\times10^3$,
$\Delta=196\pm20\,\mev$. 
\begin{figure}
  \includegraphics[height=9cm,angle=-90]{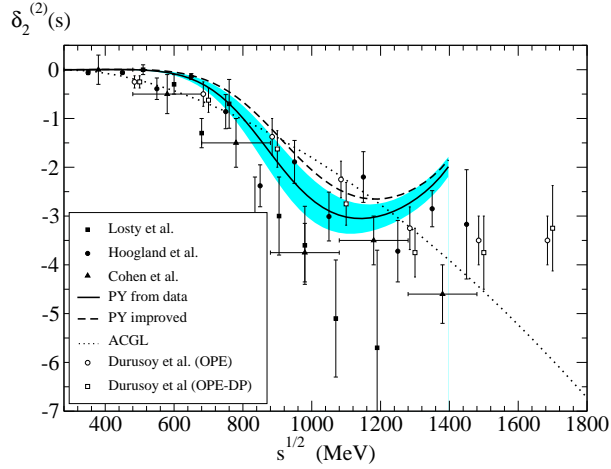} 
  \caption{Continuous line: The  
$I=2$, $D$-wave phase shift, obtained by only fitting the 
experimental data. Broken line: with the parameters improved using 
dispersion relations. Dotted line: the 
fit, valid between $s^{1/2}=0.625\,\gev$ and 1.375~\gev, of 
Martin, Morgan and Shaw which ACGL and CGL, however, use from threshold 
to  $s^{1/2}=2\,\gev$.  
The experimental points are from \cite{7}. }
\end{figure}
  The fit, which may be found in Fig~2, returns 
 reasonable
 numbers for the scattering length and  for the effective range parameter,
$b_2^{(2)}$: 
\begin{equation}
a_2^{(2)}=(2.5\pm0.9)\times10^{-4}\,{M_\pi}^{-5};\quad
b_2^{(2)}=(-2.7\pm0.8)\times10^{-4}\,{M_\pi}^{-7}.
\label{(2.5)}
\end{equation}

\section{The high energy ($s^{1/2}\geq 1.42\,\gev$) input}

\noindent
In order to test dispersion relations we also need the 
 imaginary part of the scattering amplitude at $s^{1/2}\geq1.42\;\gev$,
that we  take  
from a Regge fit to data \cite{4}
(and the slightly improved rho residue of \cite{5}).
We note that, in the early 1970s, when $\pi\pi$ phase shifts 
were  poorly known and, 
above all, when it was still not
 clear that the standard Regge picture is a QCD feature,
Regge factorization was questioned \cite{13}
using crossing sum rules 
and then-existing low energy phase shift data.
This was adopted later by ACGL and CGL, assuming 
a too large rho residue and 
 a Pomeron a {\sl third} of what factorization and
the experimental data on the total $\pi\pi$ cross section implies, 
as well as unconventional slopes. Unfortunately this has been 
also used in subsequent Roy equation analyses. 
As discussed in  \cite{4,5}, 
however, standard Regge factorization
describes experiment \cite{14,Pelaez:2004ab} and 
is perfectly consistent with
crossing  sum rules if assumed to hold
above $1.42\;\gev$.

In Fig~3 we show our Regge description 
of the imaginary parts of $\pi\pi$ scattering
amplitudes \cite{4,5,Pelaez:2004ab} together with the data\cite{14},
compared  with that used by ACGL\cite{1},
CGL\cite{2} above 2 GeV. 

Between $1.42\,\gev\leq s^{1/2}\leq2\,\gev$,
these authors
use the scattering amplitude reconstructed from 
one Cern--Munich phase shift analysis, and, in particular
for  S0, the  
re-elaboration of Au, Morgan and Pennington \cite{10}.
Unfortunately,   
in this region the inelasticity is large 
and the Cern--Munich experiments, which only measure the 
{\sl differential} cross section for 
$\pi\pi\to\pi\pi$ are insufficient to reconstruct without ambiguity 
the full imaginary part. 
In addition, in \cite{3,5} we showed that 
Cern--Munich phases fail to pass a number of consistency tests. 
This is also seen clearly in Fig~3, where we plot the total 
cross section for $\pi^+\pi^-$ 
that follows Hyams et al.,\cite{10}, which is
incompatible with other experimental data \cite{14},
as well as with Regge factorization.

\begin{figure}[h]
  \begin{minipage}{\textwidth}
    {\includegraphics[height=.22\textheight]{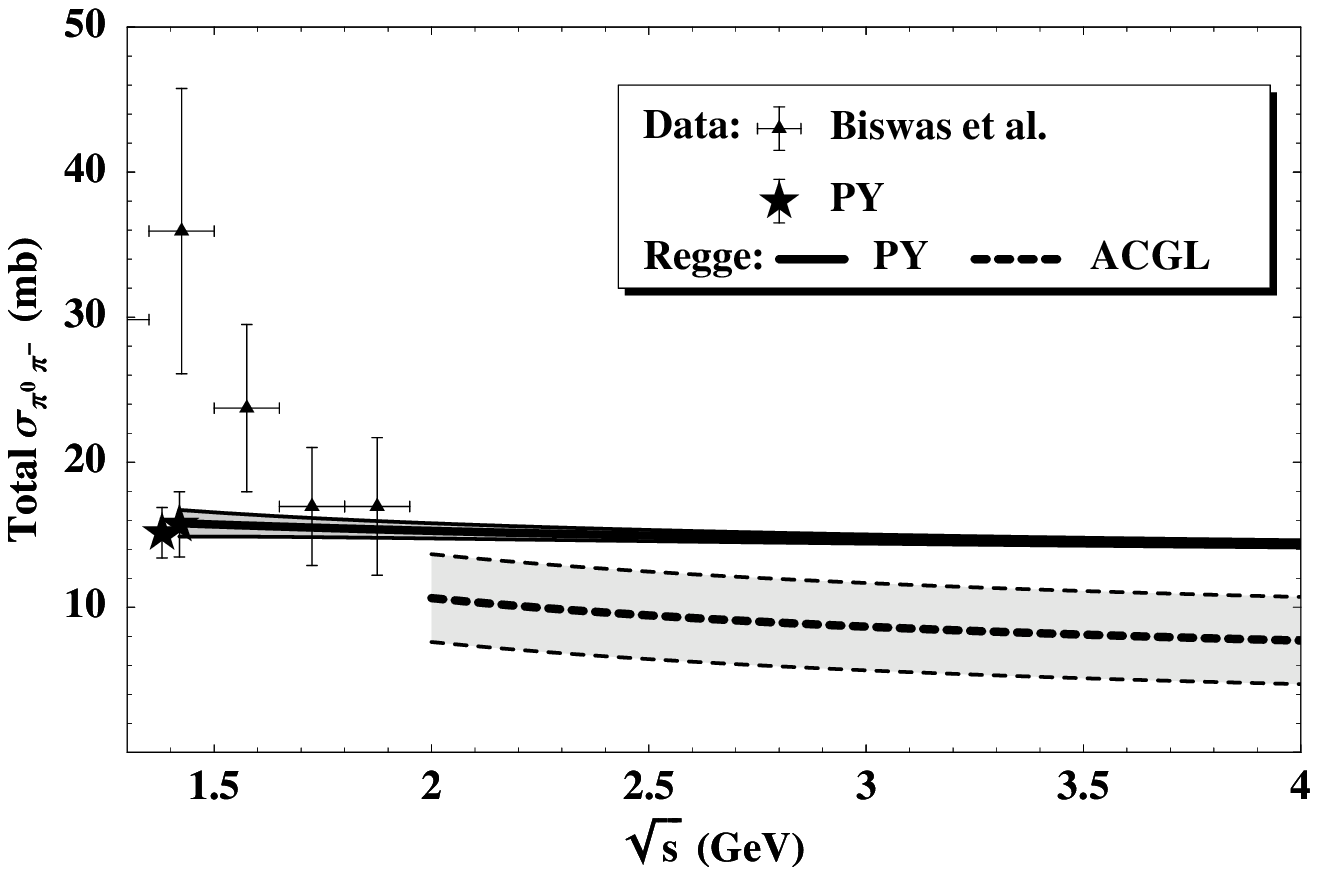}
      \includegraphics[height=.22\textheight]{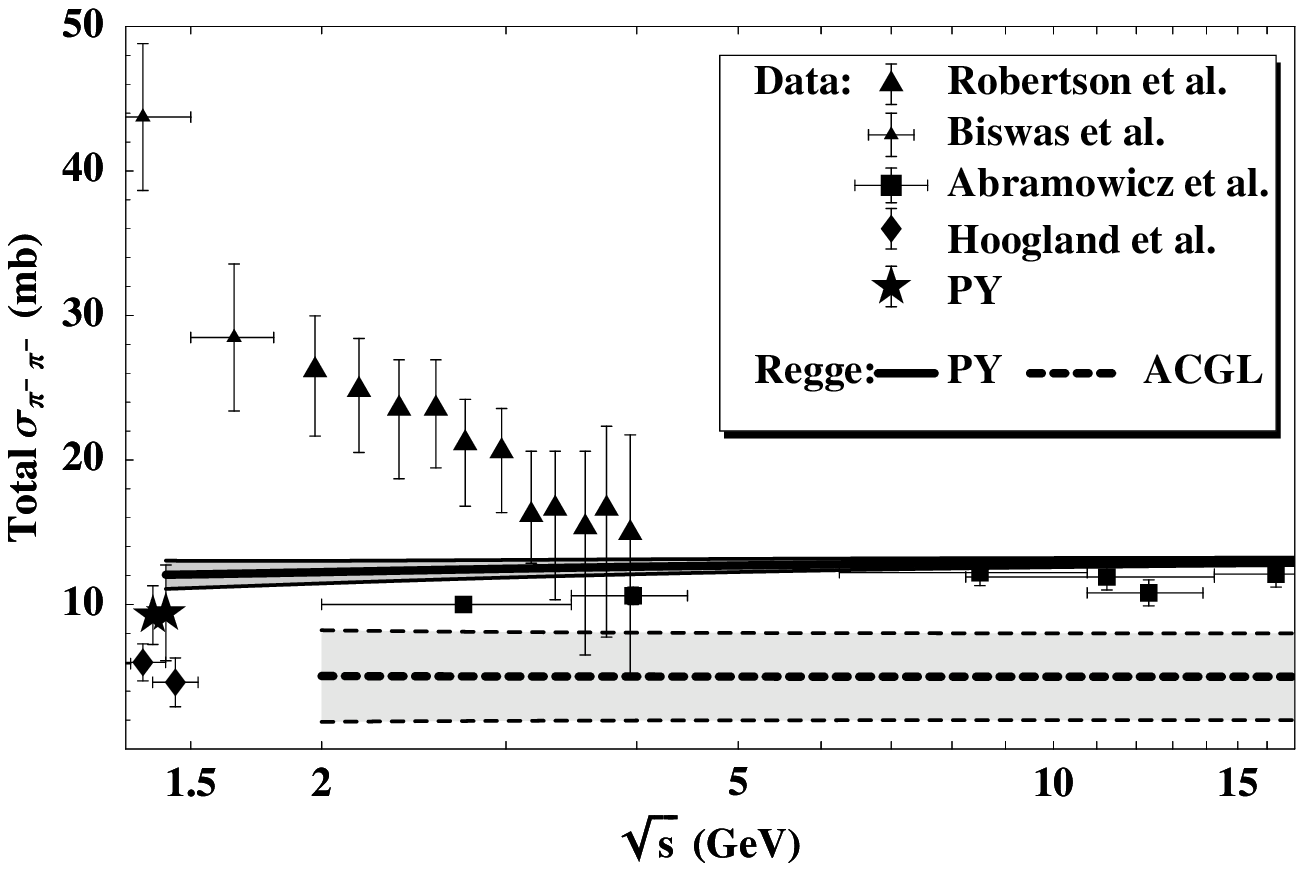}}
{ \includegraphics[height=.23\textheight]{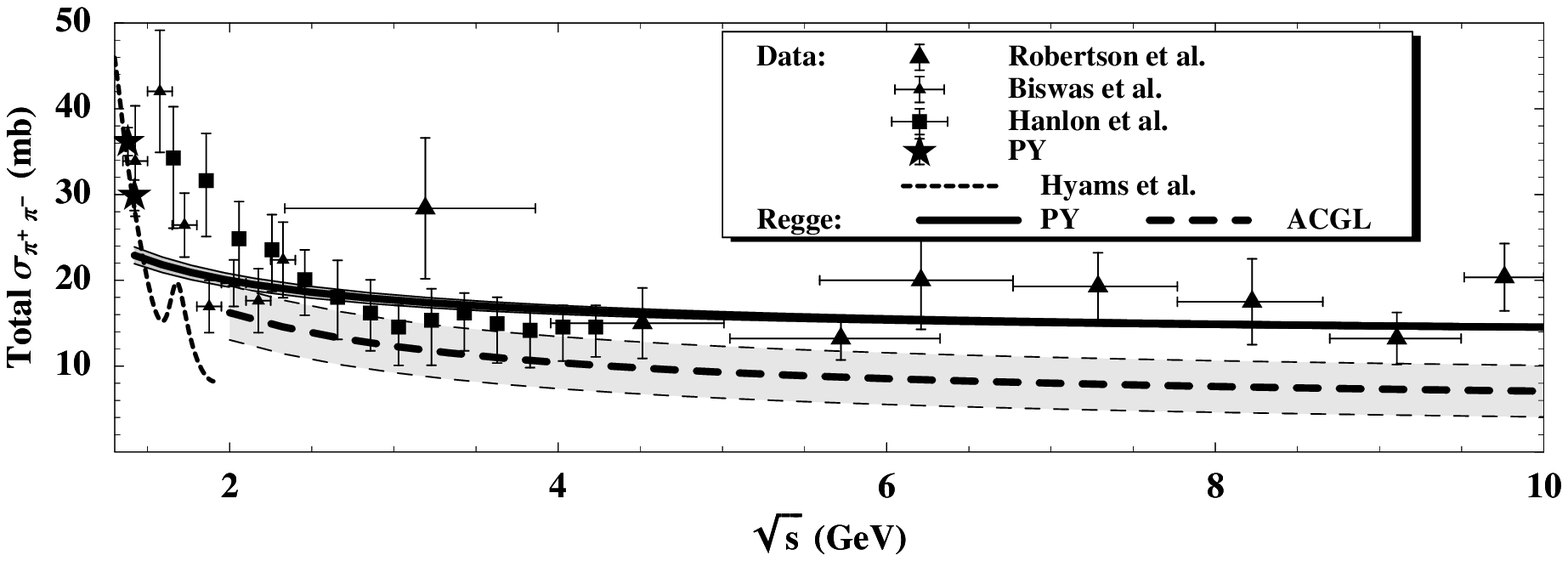}}
\end{minipage}
  \caption{The $\pi\pi$ cross sections. Experimental points from \cite{14}.
The stars at 1.38 and 1.42 \gev\ (PY) are from the phase shift analysis 
of experimental data given in \cite{5}. 
Continuous lines, from 1.42 \gev\ (PY): Regge formula, with parameters as in 
\cite{4}
(the three lines per fit  cover the error in the theoretical values of the 
Regge residues). 
Dashed lines, above 2 \gev: the cross sections following from ACGL;\cite{1} 
the gray band covers their error band.
Below 2 \gev, the dotted line 
corresponds to  the $\pi^+\pi^-$ cross section from 
 the Cern--Munich analysis; cf.~Fig~7 in the paper of Hyams~et~al.\cite{10}}
\end{figure}

 In a recent paper Caprini, Colangelo, Gasser and
Leutwyler,\cite{20}  to be denoted by CCGL, review our work in \cite{3} 
and conclude that, still, they consider the CGL solution 
consistent. They also raised the contention that our
Reggeistics could not be correct because it violates certain sum rules.
In view of Fig.3 this contention is meaningless since
the PY cross sections are perfectly compatible with
high energy ($s^{1/2}\geq1.42 \gev$) {\sl experimental} data, 
while the ACGL ones are not. In \cite{3,4,5}
we also checked that our representation satisfies two crossing sum rules.

Concerning D2, ACGL and CGL 
borrow an old fit in the book of 
Martin, Morgan and Shaw,\cite{15} where 
only intermediate energy data were fitted. 
 \begin{equation}
\delta_2^{(2)}(s)=-0.003(s/4M^2_\pi)\left(1-4M^2_\pi/s\right)^{5/2},
\label{(3.1)}
\end{equation}
which fails at threshold (it gives a negative scattering length) 
and does not fit well data below 1.42~\gev, as 
shown in Fig~2. 
Above 1~\gev, this D2 phase grows 
quadratically with the energy, while Regge theory predicts
all phases to go to a multiple of $\pi$. In particular 
D2 should go to zero; see
Appendix~C of \cite{5} for details. It is true that this D2 wave is small but, given the accuracy
claimed by CGL,  it is certainly not negligible.

\section{Checking forward dispersion relations}

In the present Section we study how well the previous
amplitudes obtained from fits to different sets of data
satisfy forward dispersion relations. 
We consider 
three independent scattering amplitudes in
$t$-symmetric or antisymmetric combinations, 
that form a complete set: 
$\pi^0\pi^0\to\pi^0\pi^0$, $\pi^0\pi^+\to\pi^0\pi^+$, and the 
$t$ channel isospin one amplitude, $I_t=1$.
The reason is that the two first
depend  only on two isospin states, and have positivity properties: 
their imaginary parts are sums of positive terms, thus reducing the
final uncertainties. Hence,  for $\pi^0\pi^0$, we have
\begin{eqnarray}
\real F_{00}(s)-F_{00}(4M_{\pi}^2)=
\dfrac{s(s-4M_{\pi}^2)}{\pi}\pepe\int_{4M_{\pi}^2}^\infty\dd s'\,
\dfrac{(2s'-4M^2_\pi)\imag F_{00}(s')}{s'(s'-s)(s'-4M_{\pi}^2)(s'+s-4M_{\pi}^2)}.
\label{(4.1a)}
\end{eqnarray}
In particular, for $s=2M^2_\pi$, which will be important for the Adler zeros, we have
\begin{equation}
F_{00}(4M_{\pi}^2)=F_{00}(2M_{\pi}^2)+D_{00},\quad D_{00}=
\dfrac{8M_{\pi}^4}{\pi}\int_{4M_{\pi}^2}^\infty\dd s\,
\dfrac{\imag F_{00}(s)}{s(s-2M_{\pi}^2)(s-4M_{\pi}^2)}.
\label{(4.1b)}
\end{equation}
For the $\pi^0\pi^+$ channel, which does not depend on S0:
$$\real F_{0+}(s)-F_{0+}(4M_{\pi}^2)=
\dfrac{s(s-4M^2_\pi)}{\pi}\pepe\int_{4M_{\pi}^2}^\infty\dd s'\,
\dfrac{(2s'-4M^2_\pi)\imag F_{0+}(s')}{s'(s'-s)(s'-4M_{\pi}^2)(s'+s-4M_{\pi}^2)}.
\label{(4.2a)}$$
At the point $s=2M^2_\pi$, this becomes
\begin{equation}
F_{0+}(4M_{\pi}^2)=F_{0+}(2M_{\pi}^2)+D_{0+},\quad
D_{0+}=\dfrac{8M_{\pi}^4}{\pi}
\int_{8M_{\pi}^2}^\infty\dd s\,\dfrac{\imag F_{0+}(s)}{s(s-2M_{\pi}^2)(s-4M_{\pi}^2)}.
\label{(4.2b)}
\end{equation}
Finally, for isospin unit exchange, which does not require subtractions, 
\begin{equation}
\real F^{(I_t=1)}(s,0)=\dfrac{2s-4M^2_\pi}{\pi}\,\pepe\int_{4M^2_\pi}^\infty\dd s'\,
\dfrac{\imag F^{(I_t=1)}(s',0)}{(s'-s)(s'+s-4M^2_\pi)}. 
\label{(4.3)}
\end{equation}
at threshold this is known as the Olsson sum rule.

Depending on the method we use to fit the S0 wave we find the results in Table~1,
where, we have separated on top those fits to data with a total
 $\chi^2/d.o.f.<6$ for the $\pi^0\pi^0$ and $I_t=1$ dispersion relations
up to 0.925 GeV, 
a fairly reasonable $\chi^2/d.o.f.$ since these
fits were obtained independently of the dispersive approach. 
\begin{table}
\begin{tabular}{|c|c|c|c|c|c|c|}
\hline
&$B_0$&$B_1$&$M_\sigma$ (MeV)&${I_t=1}\atop {\dfrac{\chi^2}{\rm d.o.f.}}$&
${\pi^0\pi^0}\atop{\dfrac{\chi^2}{\rm d.o.f.}}$&$\delta_0^{(0)}(0.8^2)$\\
\hline
PY,  Eqs.(\ref{(2.3a)},\ref{(2.3b)})&$21.04$&
$6.62$&$782\pm24$&$0.3$ &
$3.5$&
$91.9\degrees$\\\hline
$K\; {\rm decay\;  only}$
&$\phantom{\Big|}18.5\pm1.7$&$\equiv0$&$766\pm95$
&$0.2$  &$1.8$&
$93.2\degrees$ \\ \hline
${\displaystyle{{ K\; {\rm decay\; data}}}\atop{\displaystyle +\,{\rm Grayer,\;B}}}$&
$22.7\pm1.6$ &$12.3\pm3.7 $&$858\pm15 $&
$1.0$&$2.7$&
$84.0\degrees$\\
\hline
${\displaystyle{{ K\; {\rm decay\; data}}}\atop{\displaystyle +\,{\rm Grayer,\;C}}}$
&
$ 16.8\pm0.85$ & 
$-0.34\pm2.34$ & $787\pm9 $
&$0.4$&$1.0$&
$91.1\degrees$\\
\hline
${\displaystyle{{ K\; {\rm decay\; data}}}\atop{\displaystyle +\,{\rm Grayer,\;E}}}$&
$21.5\pm3.6 $&$12.5\pm7.6 $&$1084\pm110 $&
$2.1$&$0.5$&
$70.6\degrees$\\
\hline
${\displaystyle{{ K\; {\rm decay\; data}}}\atop{\displaystyle +\,{\rm Kaminski}}}$&
$ 27.5\pm3.0$&$21.5\pm7.4 $&$789\pm18 $&$0.3$&$5.0$&
$91.6\degrees$
\\\hline
\hline
${\displaystyle{{ K\; {\rm decay\; data}}}\atop{\displaystyle +\,{\rm Grayer,\;A}}}$&
$ 28.1\pm1.1$&$26.4\pm2.8 $&$866\pm6 $
&$2.0$&$7.9$&
$81.2\degrees$\\
\hline
${\displaystyle{{ K\; {\rm decay\; data}}}\atop{\displaystyle +\,{{\rm EM},\;s{\rm -channel}}}}$&
$ 29.8\pm1.3$&$25.1\pm3.3 $&$811\pm7 $&$1.0$&$9.1$&
$88.3\degrees$\\
\hline
${\displaystyle{{ K\; {\rm decay\; data}}}\atop{\displaystyle +\,{{\rm EM},\;t{\rm -channel}}}}$&
$ 29.3\pm1.4$&$26.9\pm3.4 $&$829\pm6 $&$1.2$&$10.1$&
$85.7\degrees$\\
\hline
${\displaystyle{{ K\; {\rm decay\; data}}}\atop{\displaystyle +\,{\rm Protopopescu\,VI}}}$&
$ 27.0\pm1.7$&$22.0\pm4.1 $&$855\pm10 $&$1.2$&$5.8$&
$82.9\degrees$\\
\hline
${\displaystyle{{ K\; {\rm decay\; data}}}\atop{\displaystyle +\,{\rm Protopopescu\,XII}}}$&
$ 25.5\pm1.7$&$18.5\pm4.1 $&$866\pm14 $&$1.2$&$6.3$&
$82.2\degrees$\\
\hline
${\displaystyle{{ K\; {\rm decay\; data}}}\atop{\displaystyle +\,{\rm Protopopescu\,3}}}$&
$ 27.1\pm2.3$&$23.8\pm5.0 $&$913\pm18 $&
$1.8$&$4.2$&
$76.7\degrees$\\
\hline
\end{tabular}
\caption{
 PY: our 
global fit, Eqs.(\ref{(2.3a)},\ref{(2.3b)}).  
We do not give its $B_0$ and $B_1$ uncertainties
as they are strongly correlated, see Eq.(\ref{(2.3b)}) for the uncorrelated ones.
Grayer B, C, E: different solutions in Grayer et al.\cite{10}. 
Kaminski: \cite{10}.  In \cite{5} we have also
studied fits to the data in Tables VI, XII and VIII in \cite{9},
to Solution A in\cite{10}, 
as well as fits to the theoretical outcome in Estabrooks and  Martin.\cite{10}.
They all give a total $\chi^/{\rm d.o.f.}\geq6$}
\label{tab:a1}
\end{table}

However, in Table 1 we also list the very frequently used 
$t$ and $s$-channel solutions of Estabrooks and Martin 
\cite{10}, those of Protopopescu {\it et al.}\cite{9}, 
from Table VI, VIII and table XII, as well as 
the solution A of Grayer {\it et al.} \cite{10}. 
Their  $I_t=1$ plus $\pi^0\pi^0$ dispersion relation
total $\chi^2/d.o.f.$ is surprisingly poor: 11.3, 10.1, 7, 6, 7.5, 9.9, respectively.
{\it Therefore, any result that relies heavily on these
sets should be taken very cautiously}.

\section{ Improved fits using dispersion relations}

\noindent
We now improve the previous low energy fits parameters
by fitting also the 
 dispersion relations up to 0.925~\gev, thus obtaining parametrizations
more compatible with analyticity and $s\,-\,u$ crossing. 
This is an alternative method to Roy equations;
it is better in that we do not need as input the scattering amplitude for
$|t|$ up to
$30M^2_\pi$, where the Regge fits  
existing in the literature disagree strongly
(see \cite{5}, Appendix~B)
 and also in that we can 
test all energies\cite{5}, whereas Roy equations are valid for
$s^{1/2}<\sqrt{60}\,M_\pi\sim 1.1\,\gev$
(and only applied up to $0.8\,\gev$). 
Starting from Eqs.(\ref{(2.3a)},\ref{(2.3b)}), we find, in $M_\pi$ units,
\begin{eqnarray}
{\rm S0};\; s^{1/2}\leq 2m_K:&&  B_0=17.4\pm0.5;\; B_1=4.3\pm1.4;
\cr && M_\sigma=790\pm21\,\mev;\;
z_0=195\,\mev\;\hbox{[Fixed]};\cr
&&  a_0^{(0)}=0.230\pm0.015;\; b_0^{(0)}=0.312\pm0.014.\cr
{\rm S2};\; s^{1/2}\leq 1.0:&& B_0=-80.8\pm1.7;\; B_1=-77\pm5;\;
z_2=147\,\mev\;\hbox{[Fixed]};\cr
&&  a_0^{(2)}=-0.0480\pm0.0046;\; b_0^{(2)}=-0.090\pm0.006.\cr
{\rm S2};\;  1.0\leq s^{1/2}\leq1.42:&&  B_0=-125\pm6;\; B_1=-119\pm14;\;
\epsilon=0.17\pm0.12.\cr
{\rm P};\; s^{1/2}\leq 1.05:&& 
B_0=1.064\pm0.11;\; B_1=0.170\pm0.040;\;
M_\rho=773.6\pm0.9\;\mev;\cr
&& a_1=(38.7\pm1.0)\times10^{-3};\; b_1=(4.55\pm0.21)\times10^{-3}.\cr
 {\rm D0};\; s^{1/2}\leq
1.42:&& B_0=23.5\pm0.7;\; B_1=24.8\pm1.0;\; \epsilon=0.262\pm0.030;\cr
&& \; a_2^{(0)}=(18.4\pm3.0)\times10^{-4};\; b_2^{(0)}=(-8.6\pm3.4)\times10^{-4}.
\cr  {\rm D2};\; s^{1/2}\leq 1.42:&&  
B_0=(2.9\pm0.2)\times10^3;\; B_1=(7.3\pm0.8)\times10^3;\;\cr
&&
B_2=(25.4\pm3.6)\times10^3;\;\Delta=212\pm19;\cr
&& a_2^{(2)}=(2.4\pm0.7)\times10^{-4};\; b_2^{(2)}=(-2.5\pm0.6)\times10^{-4}.\cr
{\rm F};\; s^{1/2}\leq 1.42:&& B_0=(1.09\pm0.03)\times10^5;\;
B_1=(1.41\pm0.04)\times10^5;\nonumber\\
&&a_3=(7.0\pm0.8)\times10^{-5}.
\label{(4.4)}
\end{eqnarray}
In Fig.4 we show the improved curves for S0 and S2, 
and  that of D2 in Fig.2.
\begin{figure}[h]
\includegraphics[height=.38\textheight,angle=-90]{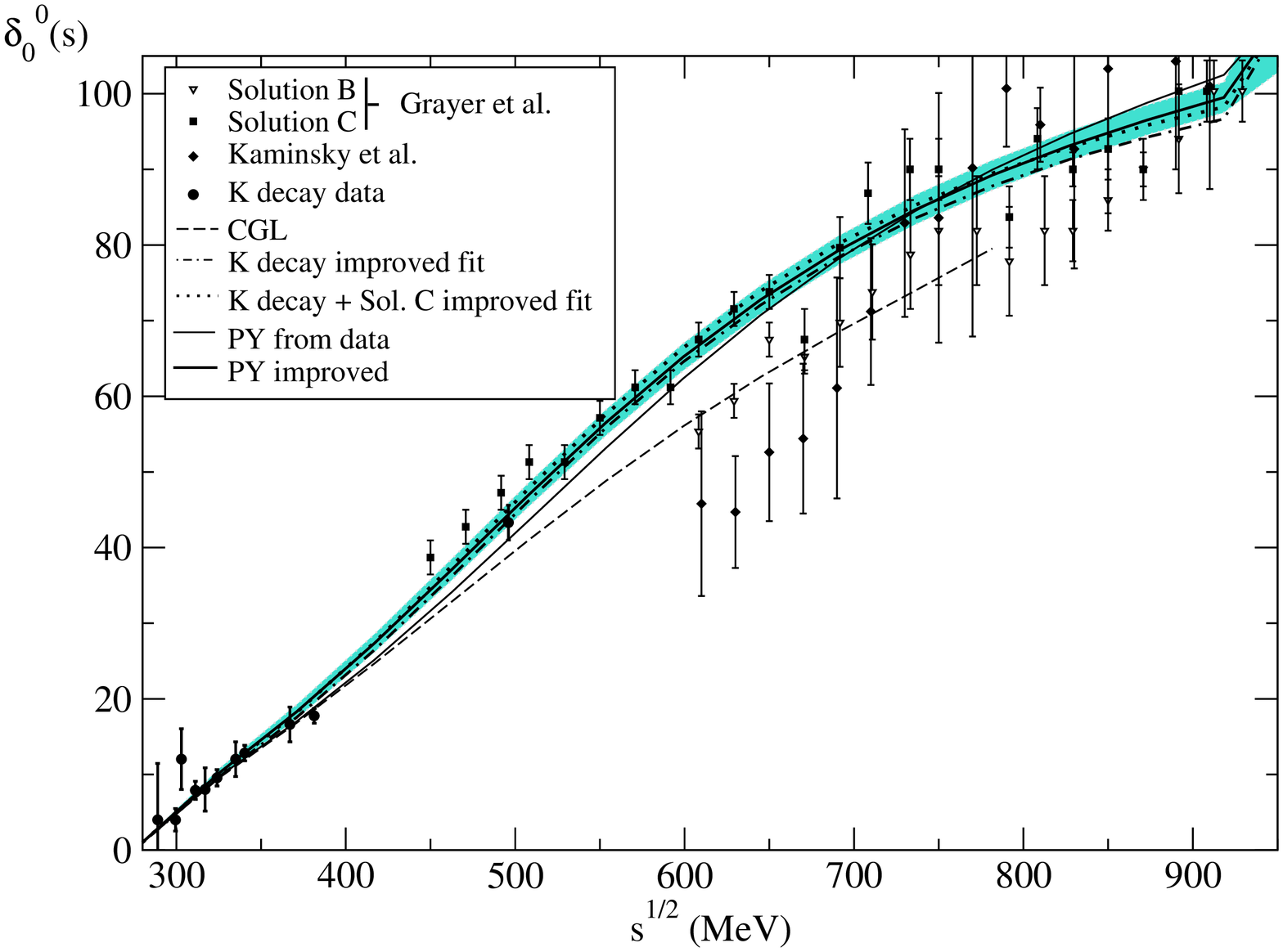}
\hspace*{-1.1cm}
\includegraphics[height=.38\textheight,angle=-90]{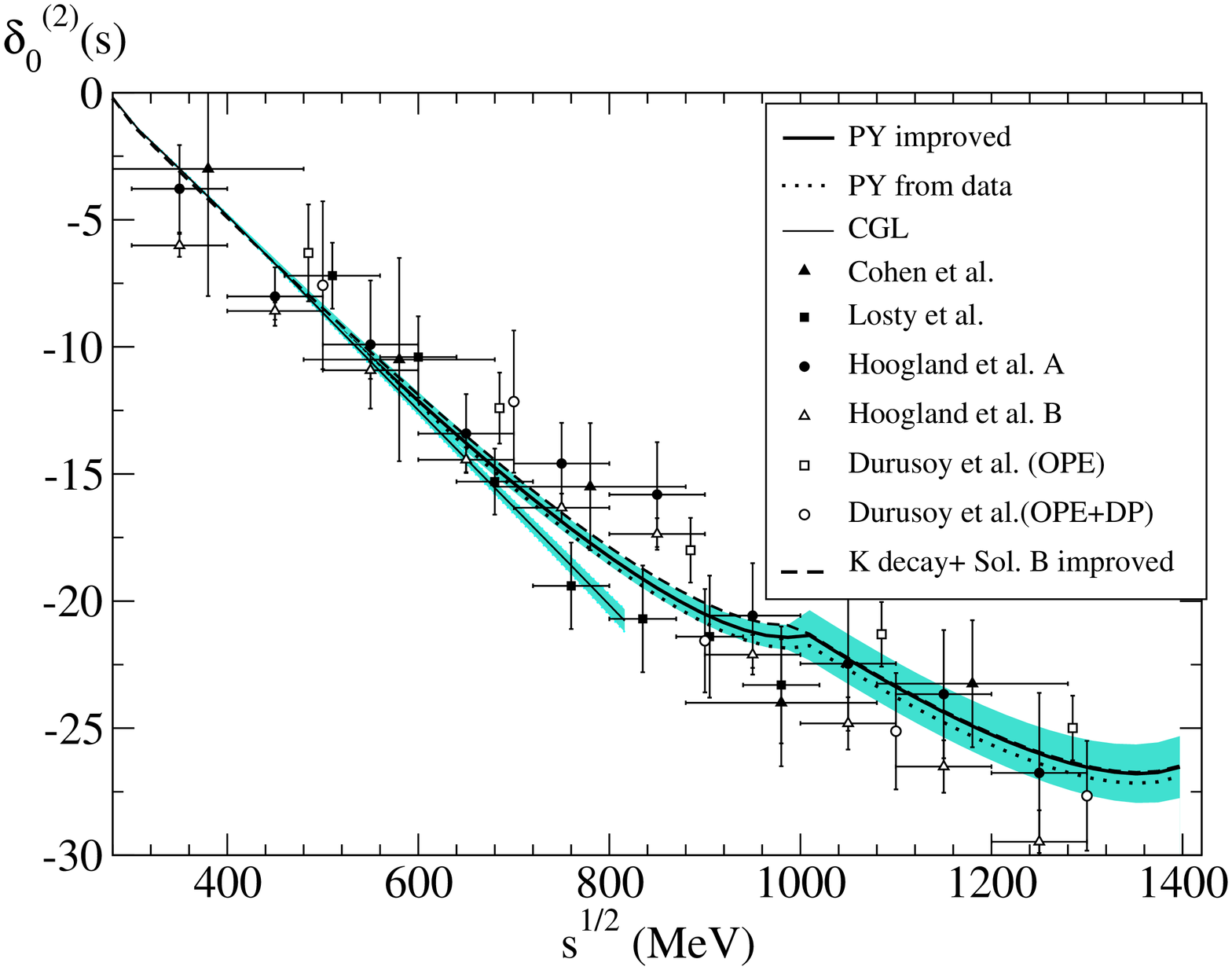}
  \caption{a) The improved S0 phase shift (PY improved, Eq.\ref{(4.4)}), 
the global fit (PY from data, he S0 
Eqs.~(\ref{(2.3a)},\ref{(2.3b)})),
 and the 
{\sl improved} solutions ``$K$ decay only" and
 ``Grayer~C" of Table~2 (almost on top of PY improved). 
The solution CGL\cite{2} (dashed line) is also shown. 
b) S2 improved Phase shift (PY improved,   
  Eq.~(\ref{(4.4)})); global fit (PY from data,  Eq.~(\ref{(2.2b)}));
 the solution CGL \cite{2} (thin continuous line) and the improved
parametrization with K decays and So. B of Grayer et al.\cite{10}.}
\end{figure}

Concerning the improved fits to individual sets of data, we get 
somewhat different results for S0, listed in Table 2.
In Table 2 we also show
the $\chi^2/d.o.f.$ of each forward dispersion relation and the
standard deviations for the sum rule in Eq.\ref{(4.1b)}, which are more than four
for  K decay plus the Grayer B or E 
or Kaminski improved solutions. 
Concerning the other waves, no matter what
set of parameters from data fits we start from,
we end up with very similar values to those given in Eq.\ref{(4.4)}.
This can be checked in  Fig.4.b, where we show
the improved ``K decay + Grayer Sol. B'' S2 wave. Even though
it is the one for which we obtained the 
most different central values for the S0 wave compared with those
given in Eq.\ref{(4.4)}, it falls perfectly within the uncertainty
of our improved solution. 

\begin{table}[htbp]
  \centering
  \caption{
Improved fits. Names are as in Table~1.
Although errors are given for the Adler zero, we fix it 
when evaluating other errors, to break the otherwise very large 
correlations}
\begin{tabular}{|c|c|c|c|c|c|c|}
\hline
Improved& Improved  &$K$ decay only& $K$ decay &
$K$ decay &$K$ decay &$K$ decay \\
fits:&PY, Eq.\ref{(4.4)}&&+grayer C& +Grayer B&+Grayer E&+ Kami\'nski\\
\hline
$B_0$ & $ 17.4\pm0.5$&$16.4\pm0.9$&$16.2\pm0.7$ &$ 20.7\pm1.0$
&$ 20.2\pm2.2$&$ 20.8\pm1.4$\\
$B_1$&$4.3\pm1.4$&$\equiv0$&$0.5\pm1.8 $&$11.6\pm2.6 $&$8.4\pm5.2 $
&$13.6\pm43.7 $\\
$M_\sigma\,$(MeV)&$790\pm30$&$809\pm53$&$788\pm9 $
&$861\pm14 $&$982\pm95 $&$798\pm17$ \\
$z_0\,$(MeV)&$195\pm30$&$182\pm34$&$182\pm39 $&$233\pm30 $
&$272\pm50 $&$245\pm39 $\\
\hline
${\displaystyle I_t=1}\atop{\chi^2/ d.o.f.}$&0.40&0.30&0.37&0.37&0.60&0.43\\
\hline
${\displaystyle \pi^0\pi^0}\atop{\chi^2/ d.o.f.}$&0.66
&0.29&0.32&0.83&0.09&1.08\\
\hline
${\displaystyle \pi^+\pi^-}\atop{\chi^2/ d.o.f.}$&
1.62&1.77&1.74&1.60&1.40&1.36\\
\hline
Eq.(\ref{(4.1b)})&1.6$\sigma$&1.5$\sigma$&1.5$\sigma$&4.0$\sigma$&6.0$\sigma$&
4.5$\sigma$\\
\hline
$\delta^{(0)}_0(0.8^2 {\rm GeV}^2)$& 91.3\degrees&
91.3\degrees&91.0\degrees&85.1\degrees&78.0\degrees&91.8\degrees\\
\hline
\end{tabular}
  \label{tab:2}
\end{table}

\section{Dispersion relations and the CGL solution}

We have also checked the fulfillment
of forward dispersion relations for the 
CGL solution for the S0, S2 and P waves al low energy. 
This is depicted in Fig~5, 
where we show, both for CGL and our improved fit, Eq.(\ref{(4.4)}),
 the mismatch between the real part and the
dispersive 
evaluations, that is to say, the differences $\Delta_i$, 
\begin{eqnarray}
&&\Delta_{1}\equiv\real
F^{(I_t=1)}(s,0)-\dfrac{2s-4M^2_\pi}{\pi}\pepe\int_{4M^2_\pi}^\infty\dd s'\,
\dfrac{\imag F^{(I_t=1)}(s',0)}{(s'-s)(s'+s-4M^2_\pi)},
\label{(7.1a)}\\
&&\Delta_{00}\equiv\real F_{00}(s)-F_{00}(4M_{\pi}^2)\\&&\qquad\quad-
\dfrac{s(s-4M_{\pi}^2)}{\pi}\pepe\int_{4M_{\pi}^2}^\infty\dd s'\,
\dfrac{(2s'-4M^2_\pi)\imag F_{00}(s')}{s'(s'-s)(s'-4M_{\pi}^2)(s'+s-4M_{\pi}^2)},
\label{(7.1b)}\nonumber\\
&&\Delta_{0+}\equiv\real F_{0+}(s)-F_{0+}(4M_{\pi}^2)\\&&\qquad\quad-
\dfrac{s(s-4M^2_\pi)}{\pi}\pepe\int_{4M_{\pi}^2}^\infty\dd s'\,
\dfrac{(2s'-4M^2_\pi)\imag F_{0+}(s')}{s'(s'-s)(s'-4M_{\pi}^2)(s'+s-4M_{\pi}^2)}.
\label{(7.1c)}\nonumber
\end{eqnarray}
These quantities would vanish,  $\Delta_i=0$, 
if the dispersion relations were exactly satisfied.
 
\begin{figure}[h]
\includegraphics[height=.38\textheight]{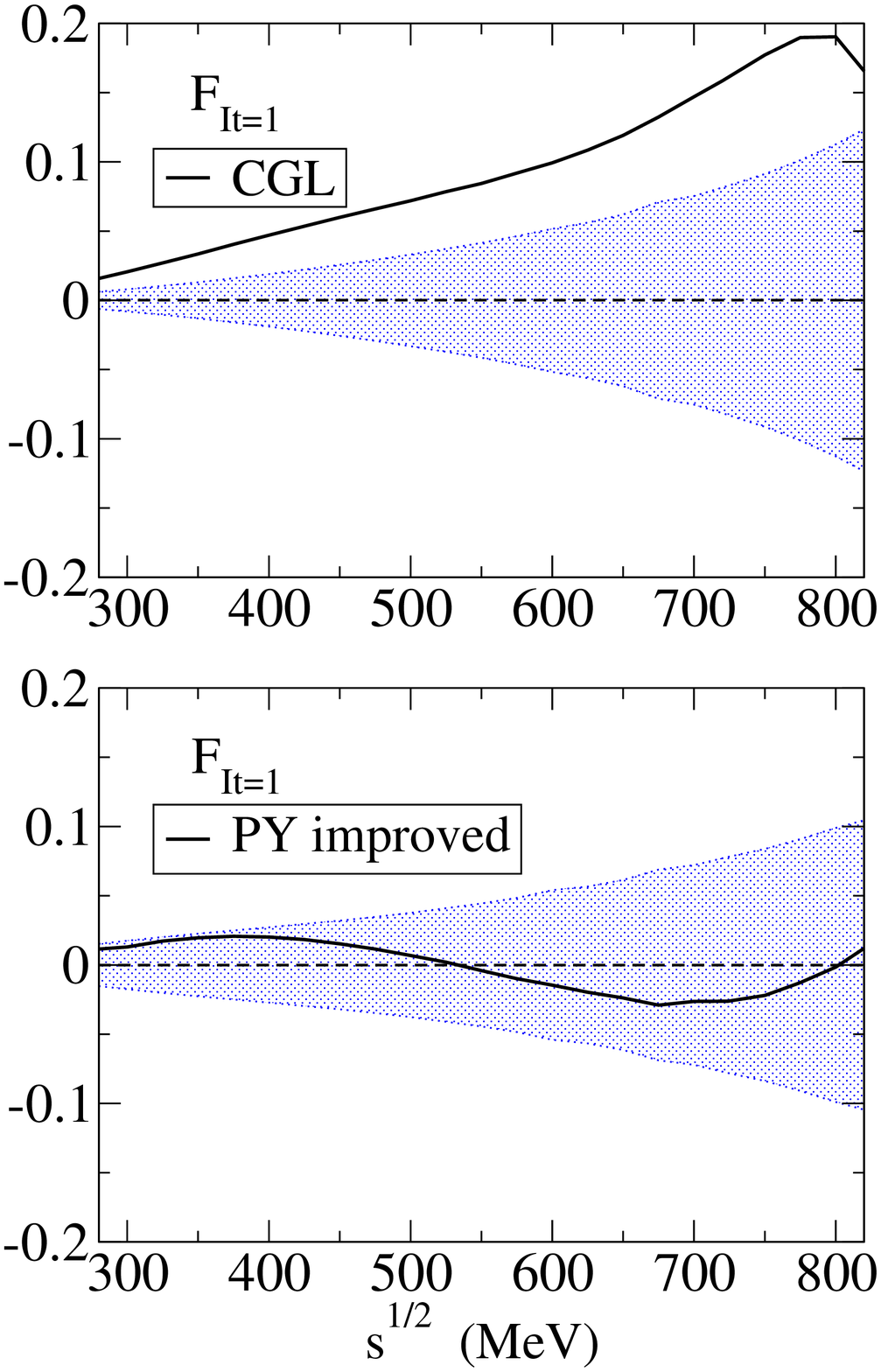}
\hspace*{-1.cm}
\includegraphics[height=.38\textheight]{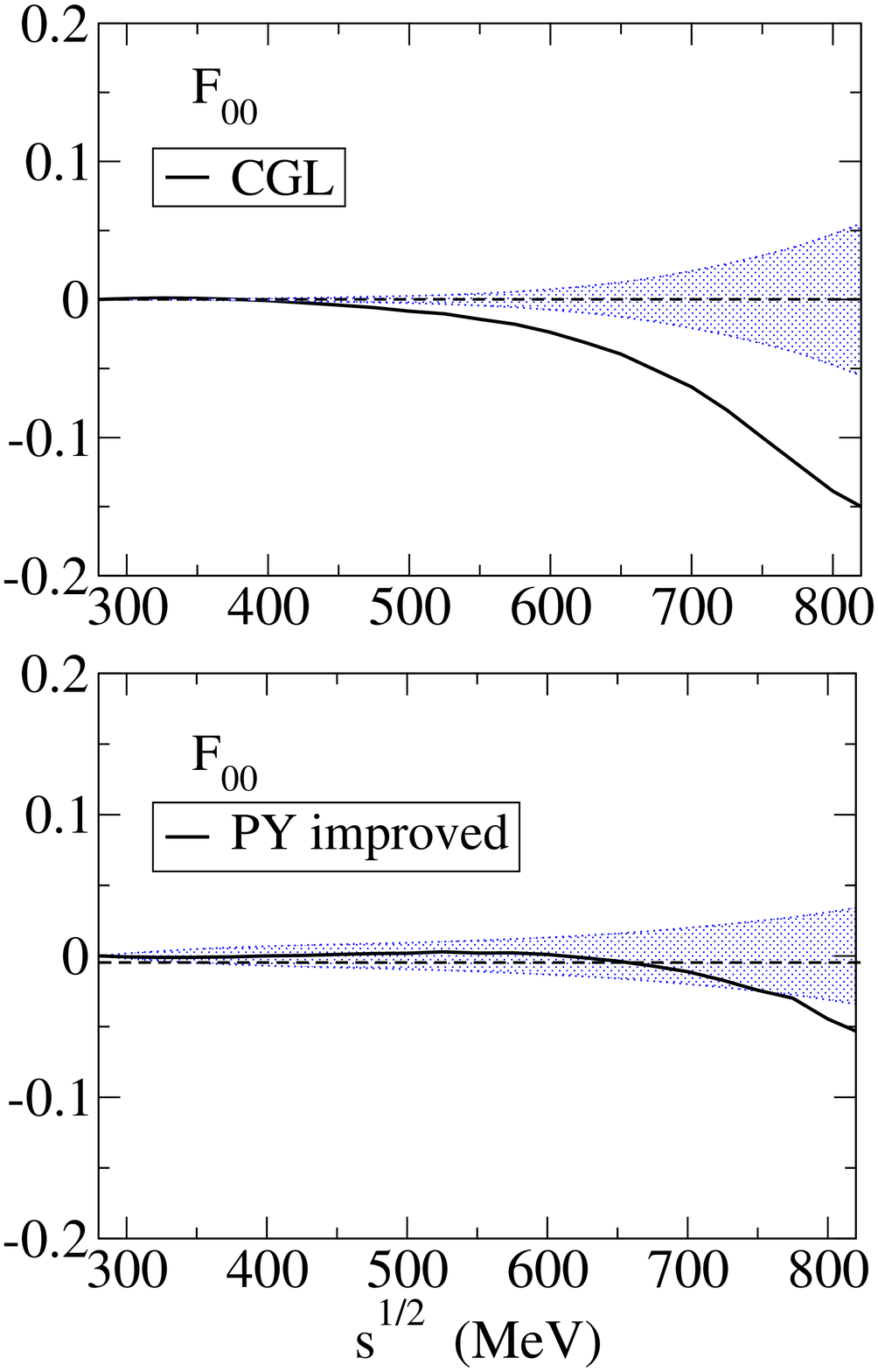}
\hspace*{-1.cm}
\includegraphics[height=.38\textheight]{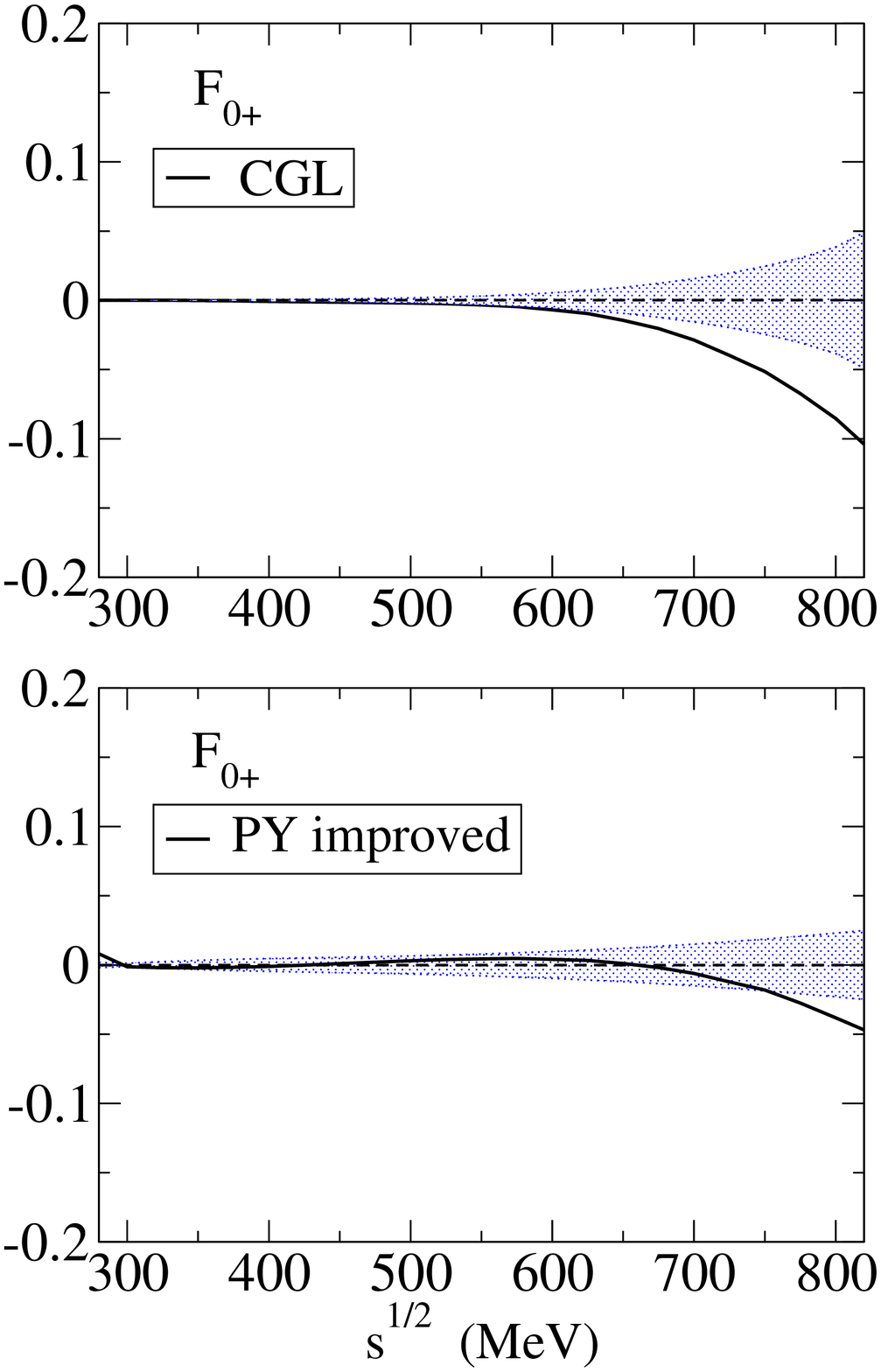}
  \caption{Dispersion
relations for the 
$\pi\pi$ amplitudes 
of \cite{2}~(CGL) and for our improved global fit
(PY improved, Eq.\ref{(4.4)}.  We plot the differences $\Delta_i$, 
Eqs.~\ref{(7.1a)}, between real parts calculated directly 
from the parametrizations, or from the 
dispersive formulas. Consistency within one sigma 
occurs within the shaded bands.
The progressive deterioration of the CGL results as the energy
 increases is apparent
here.}
\end{figure}
We  include in the comparison of Fig~5 the uncertainties; in the case of CGL, 
these errors are as follow from the parametrizations given by these authors in \cite{2}, 
for $s^{1/2}\lsim0.8\,\gev$. 
At higher energies they are taken from data via our 
parametrizations. It can be clearly seen that the CGL
parametrizations do not satisfy these forward dispersion relations by several standard deviations.

One might wonder why in Table 2,
the S0 wave improved ``K decay+solution B''
yields $\chi^2/d.o.f.$ of order one for the forward 
dispersion relations,  being so similar to what CGL get
for that wave. The reason is that, as we show in Fig.4.b. the 
improved solution B, requires an S2 wave that
is even farther from the CGL S2 wave than our global improved solution.

\section{ Low energy parameters in the literature}

\noindent
We here present, in Table~3,  the low energy parameters 
obtained from Roy equations by CGL, by Descotes et al.\cite{15}, that we denote by
DFGS, and by Kami\'nski, Le\'sniak and
Loiseau\cite{15}, denoted by KLL. 
This is compared with what we find fitting experimental data,
 improved with dispersion relations (see \cite{5} for details).

The mismatches between many of the parameters of CGL and PY are apparent 
here; not surprisingly, they
affect mostly parameters sensitive to high energy:
\begin{itemize}
\item The S wave parameters are compatible, mainly due to
  the large PY uncertainties.  
\item The mismatch between CGL and PY for $a_2^{(I)}$ and $b_2^{(I)}$
is roughly $2.5\sigma$.
In ref.\cite{3} we pointed out
  that the ACGL and CGL results did not satisfy
  the Froissart Gribov sum rules.  This happens to more than four
  standard deviations for the difference between the CGL calculation
  using Wanders sum rules minus the Froissart-Gribov representation.
  This larger mismatch, as pointed out in
 \cite{20}, does not involve the S and
  P waves, and is due to the Regge and $L\geq2$
  wave input, evidently very different from the beginning for CGL and PY,
but that certainly affects the values of  $a_2^{(I)}$ and $b_2^{(I)}$.

\item The $b_1$ calculation differs
  by more than 4 standard deviations for PY and CGL.
\end{itemize}

\begin{table}
{\footnotesize
\begin{tabular}{|c|c|c|c|c|}
\hline
& DFGS &KLL& CGL& PY\\ \hline
$a_0^{(0)}$&$0.228\pm0.032$&$0.224\pm0.013$
&$0.220\pm0.005$ &
$0.230\pm0.015$\\ 
\hline
$a_0^{(2)}$&$-0.0382\pm0.0038$&$-0.0343\pm0.0036$&
$-0.0444\pm0.0010$ &
$-0.0480\pm0.0046$\\
\hline
$b_0^{(0)}$&&$0.252\pm0.011$
&$0.280\pm0.001$ &
$0.312\pm0.014$\\
\hline
$b_0^{(2)}$&&$-0.075\pm0.015$
&$-0.080\pm0.001$ &
$-0.090\pm0.006$\\
\hline
$a_1\times10^3$&&$39.6\pm2.4$
&${\displaystyle 37.9\pm0.5}$ &
$38.4\pm0.8$
%$\;(\times\,10^{-3})$
\\
\hline
$b_1\times10^3$&&$2.83\pm0.67$&
$5.67\pm0.13$  
& 
$4.75\pm0.16$
%$\;(\times\,10^{-3})$
\\
\hline
$a_2^{(0)}\times10^4$&&&$17.5\pm0.3$&$18.70\pm0.41$
%$\;(\times\,10^{-4})$
\\
\hline
$a_2^{(2)}\times10^4$&&&
$1.70\pm0.13$&$2.78\pm0.37$
%$\;(\times\,10^{-4})$
\\
\hline
$b_2^{(0)}\times10^4$
&&&$-3.55\pm0.14$&$-4.16\pm0.30$
%$\;(\times\,10^{-4})$
\\
\hline
$b_2^{(2)}\times10^4$&&&
$-3.26\pm0.12$&$-3.89\pm0.28$
%$\;(\times\,10^{-4})$
\\
\hline
$a_3\times10^5$&&&$5.6\pm0.2$&$6.3\pm0.4$
%$\;(\times\,10^{-5})$
\\
\hline
\end{tabular}
}
\caption{Units of $M_\pi$. The numbers  in the CGL column are   as given by 
CGL in  Table~2 and elsewhere in their text. 
In PY, the values for the D, F waves parameters are from the 
Froissart--Gribov representation. 
The rest are from the fits, improved with dispersion relations, except for $a_1$ and
$b_1$ that have been taken as in \cite{5}.}
\label{tab:a}
\end{table}

\section{The ACGL, CGL phase at $s^{1/2}=0.8\,$GeV}
In the ACGL, CGL analyses, by the input 
phases for the S0, S2 and P waves at the point, $s^{1/2}=0.8\,\gev$, 
where they match the solutions to the Roy equations to 
the experimental amplitude. 
Indeed it is dominant for their Olsson sum rule
calculation, which involves the $I_t=1$ channel.

The quantity $\delta_0^{(0)}((0.8\,\gev)^2)$ is in fact given
 in  Eq.~(7.3) of ACGL as
 \begin{equation}
\delta_0^{(0)}((0.8\,\gev)^2)=82.3\pm3.4\degrees.
\label{(5.1)}
\end{equation}
whose error
may be contrasted with the estimates of \cite{5}, 
which  vary, for 
the data above 0.8 \gev, between 6\degrees\ and 18\degrees,
or with the $\delta_0^{(0)}((0.8\,\gev)^2)$ values we obtain from fits to different sets of data 
in Table 1, or the improved fits in Table 2. 
Small errors could be expected from theoretical analysis including
many data but the above small error
was used as an {\it input}.

The reason to consider such an small error is that
ACGL consider the difference $\delta_1-\delta_0^{(0)}$ at 0.8 \gev,  
in the hope that some of the uncertainties will cancel. 
Then they interpolate and then average the points from a choice
of three analysis of the CERN/Munich experiment\cite{10}
$$\delta_1((0.8\,\gev)^2)-\delta_0^{(0)}((0.8\,\gev)^2)=\cases{
23.4\pm4.0\degrees\;\hbox{[Hyams et al.]} \cr
24.8\pm3.8\degrees\;\hbox{[Estabrooks and Martin, $s$-channel]} \cr
30.3\pm3.4\degrees\;\hbox{[Estabrooks and Martin, $t$-channel]}. \cr
}
$$
and set $$\delta_0^{(0)}((0.8\,\gev)^2)-\delta_1((0.8\,\gev)^2)=26.6\pm2.8\degrees.$$
However, this error does not include systematics.
All numbers here stem from the {\sl same experiment}, 
and differ only on the method of analysis. 
Their spread is an indication of the {\sl systematic} uncertainties,
roughly an additional $\pm4\degrees$.
 In addition,
the Hyams et al. value above 
 is only one of {\sl five} 
solutions in Grayer et al.\cite{10}, and considering also
 data of Protopopescu et al.,\cite{9} 
 the systematic error  would increase to $10\degrees$.
Remarkably, Estabrooks and Martin themselves, point
out in their section 4 (first paragraph) that different
D wave input ``lead to systematic changes in $\delta_S^0$ of the order
of 10\degrees''.

%%
%% BACKMATTER
%%%%%%%%%%%%%%%%%%%%%%%%%%%%%%%%%%%%%%%%%%%%%%%%

\begin{theacknowledgments}
F.J. Yndur\'ain is grateful to the organizing committee for the opprtunity
to talk at the meeting and for financial support. 
\end{theacknowledgments}

%%%%%%%%%%%%%%%%%%%%%%%%%%%%%%%%%%%%%%%%%%%%%%%%
%% You may have to change the BibTeX style below, depending on your
%% setup or preferences.
%%
%% If the bibliography is produced without BibTeX comment out the
%% following lines and see the aipguide.pdf for further information.
%%
%% For The AIP proceedings layouts use either
%%%%%%%%%%%%%%%%%%%%%%%%%%%%%%%%%%%%%%%%%%%%

\end{document}